\documentclass[conference]{IEEEtran}
\IEEEoverridecommandlockouts
\usepackage{cite}
\usepackage{amsmath,amssymb,amsfonts}
\usepackage{algorithmic}
\usepackage{graphicx}
\usepackage{textcomp}
\usepackage{xcolor}
\def\BibTeX{{\rm B\kern-.05em{\sc i\kern-.025em b}\kern-.08em
    T\kern-.1667em\lower.7ex\hbox{E}\kern-.125emX}}
\begin{document}

\title{Adaptive Few-Shot Learning PoC Ultrasound COVID-19 Diagnostic System\\
\thanks{Identify applicable funding agency here. If none, delete this.}
}

\author{\IEEEauthorblockN{1\textsuperscript{st} Michael Karnes}
\IEEEauthorblockA{\textit{
Department of Civil, Environmental, and Geodetic Engineering} \\
\textit{The Ohio State University}\\
Columbus OH, USA \\
karnes.30@osu.edu}
\and
\IEEEauthorblockN{2\textsuperscript{nd} Shehan Perera}
\IEEEauthorblockA{\textit{Department of Electrical and Computer Engineering} \\
\textit{The Ohio State University}\\
Columbus OH, USA \\
perera.27@osu.edu}
\and
\IEEEauthorblockN{3\textsuperscript{rd} Srikar Adhikari}
\IEEEauthorblockA{\textit{Department of Emergency Medicine} \\
\textit{The University of Arizona Medical Center}\\
Tucson AZ, USA \\
sadhikari@aemrc.arizona.edu}
\and
\IEEEauthorblockN{4\textsuperscript{th} Alper Yilmaz}
\IEEEauthorblockA{\textit{Department of Civil, Environmental, and Geodetic Engineering} \\
\textit{The Ohio State University}\\
Columbus OH, USA \\
yilmaz.15@osu.edu}

}

\maketitle

\begin{abstract}
This paper presents a novel ultrasound imaging point-of-care (PoC) COVID-19 diagnostic system. The adaptive visual diagnostics utilize few-shot learning (FSL) to generate encoded disease state models that are stored and classified using a dictionary of knowns. The novel vocabulary based feature processing of the pipeline adapts the knowledge of a pretrained deep neural network to compress the ultrasound images into discrimative descriptions. The computational efficiency of the FSL approach enables high diagnostic deep learning performance in PoC settings, where training data is limited and the annotation process is not strictly controlled. The algorithm performance is evaluated on the open source COVID-19 POCUS Dataset to validate the system's ability to distinguish COVID-19, pneumonia, and healthy disease states. The results of the empirical analyses demonstrate the appropriate efficiency and accuracy for scalable PoC use. The code for this work will be made publicly available on GitHub upon acceptance.
\end{abstract}


\section{Introduction}
Coronavirus disease 2019 (COVID-19), caused by the severe acute respiratory syndrome coronavirus 2 (SARS-CoV-2), has rapidly become a global health emergency \cite{sharma_severe_2020}. As of December 2020, the virus has spread to every country infecting more than 69 million people and resulting in 1.5 million deaths worldwide \cite{COVID-19_numbers}. Insufficient medical resources have become a major challenge, especially in low-income countries. There is a critical need for fast, accessible and low-cost diagnostic tests in point-of-care (PoC) settings to stratify risk and efficiently allocate limited healthcare resources. 

SARS-CoV-2 reverse transcriptase–polymerase chain reaction (RT-PCR) is the current diagnostic gold standard worldwide \cite{oliveira_sars-cov-2_nodate}. It has an estimated sensitivity of 75\% and can take several days to obtain results \cite{woloshin_false_2020,feng_case_2020}. While chest X-ray is more widely available, its utility is limited by its low sensitivity \cite{peyrony_accuracy_2020}. Computed tomography (CT) has been viewed as an alternative for the diagnosis of COVID-19 \cite{li_diagnostic_2020}. However, due CT's ionizing radiation and limited availability outside of large hospitals it is not an optimal screening tool. A rapid, accurate, and inexpensive screening tool is required for appropriate triage and diagnosis of patients with suspected COVID-19.

The use of bedside lung ultrasound (LUS) is a common practice in a wide variety of clinical settings, including emergency departments and intensive care units \cite{staub_lung_2019}. There have been several studies evaluating the use of LUS in patients with suspected COVID-19 infection with one reporting a sensitivity of 90\%, higher than seen in X-ray \cite{haak_diagnostic_2020,convissar_application_2020}. LUS can help identify patients with COVID-19, who have been incorrectly diagnosed by RT-PCR, and prevent the further spread. In addition to screening, LUS has been shown to be an effective imaging modality for predicting the course, stratifying the risk, and monitoring COVID-19 disease state \cite{marggrander_lung_2020}. The characteristic LUS findings of COVID-19 (a thickened or irregular pleural line, confluent B-lines, sub-pleural consolidations, and pleural effusions) demonstrate promise in trending clinical progression from onset to resolution \cite{noauthor_acr_nodate}. Thus, LUS is a reliable, cost-effective, and easy-to-use tool for rapid triage, diagnosis, and early risk stratification of COVID-19.

The primary limitation of LUS diagnostics is the extensive training, experience and expertise required for the accurate identification of disease characteristics \cite{LUS_skill,Stasi_Ruoti_2015}. The ability to accurately interpret LUS images requires recognition of normal sonographic anatomy, normal variants, as well as pathology. As a result, LUS diagnostics are limited from their full use in point-of-care (PoC) settings. This creates a need for additional technologies to aid healthcare providers in interpreting LUS images. Machine learning (ML) algorithms are one such technology.

Originating from the field of pattern recognition, ML provides the framework for extracting coherent patterns from high dimensional noisy data. This is especially true for deep neural networks (DNN). In 2012, the power of the DNN was established with the record breaking performance of the AlexNet achieving a top-1 accuracy of 63.3\% over a 1000 class classification problem \cite{alexnet}. This breakthrough was made possible by the collection of a large annotated dataset, ImageNet, with millions of images and the developments in computational power. Shortly after, the DNN became larger and more complex, improving their performance each year with a current record of 88.6\% top-1 accuracy \cite{foret2020sharpnessaware}. These results solidified the position of DNN as a top approach for visual classification.

The high performances of DNN visual classification comes with a critical caveat; the training set must sufficiently represent the scenarios seen while testing. This means large annotated training sets, limited applicability, and unpredictable errors \cite{onepixel}. In response there has been an effort to reduce training set sizes and improve generalization by transferring the knowledge of a DNN pretrained on large dataset to novel applications, commonly referred to as transfer learning \cite{transfer_learning_review}. The primary advantage to this approach is the parameters of the DNN are frozen while adaptive layers are trained, which significantly reduces the number of trained parameters and the number of required training examples. This is important for applying ML to ultrasound datasets. The availability of annotated ultrasound datasets is increasing with some reaching tens of thousands of images. However, the majority of ultrasound datasets are less than 300 images \cite{2018_US_ML_Review}.

The contribution of this work is the introduction of a novel LUS diagnostic system built on the few-shot learning (FSL) visual classification algorithm. The proposed system has low training requirements with as few as 8 images per class, while traditional DNN approaches require thousands. The results of this study demonstrate the ability of the FSL based system in extending the accessibility of rapid LUS diagnostics to resource limited clinics.

\section{Related Work}
The proposed ultrasound COVID-19 diagnostic system is based on a SOTA DNN visual classification algorithm that significantly reduces training requirements associated with traditional deep learning. DNN are large regressed embedding transforms. In the classification problem, the DNN is trained to project the image space to the latent space that minimizes classification error. The learned transform filters within the network posses the network's knowledge domain. Many approaches have been taken to adapt the learned knowledge domain to novel tasks. This can take the form of fine tuning \cite{medfinetune2016,sung2018learning,snell2017prototypical}, where the learned state is used to initialize the DNN and only the final layers of network are trained \cite{Rosenfeld2020}, or direct transformations of the DNN knowledge domain to a novel task \cite{bateni2020improved,Requeima19}. The algorithm within the proposed system falls in the domain adaptation category producing a direct transform of the DNN latent space to a targeted discriminative feature space.

\begin{figure}
\begin{center}
%\fbox{\rule{0pt}{2in} \rule{0.7\linewidth}{0pt}}
\fbox{\includegraphics[width=.9\linewidth]{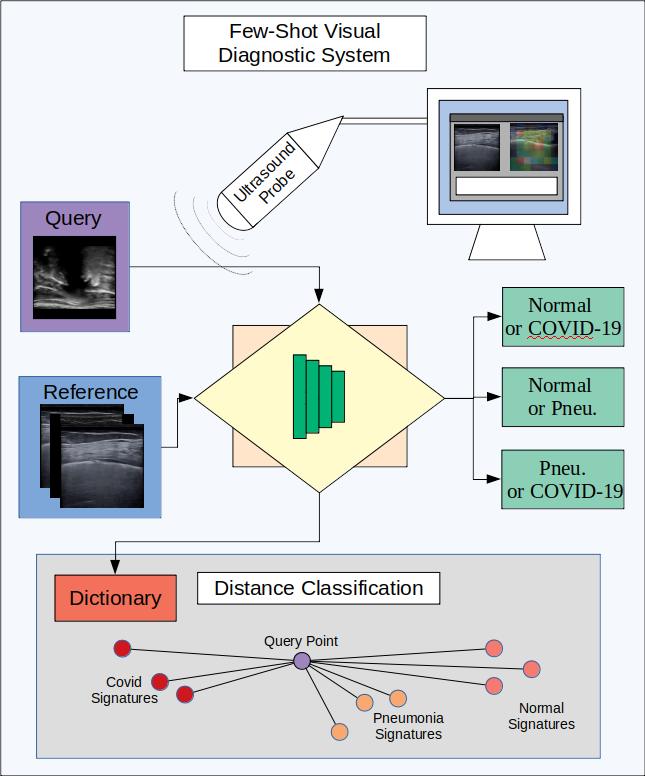}}
\end{center}
  \caption{\textbf{Diagram of Few-Shot Visual Classification:} The few-shot visual diagnostic task classifies a query image with respect to a small annotated set reference images. The ultrasound images are imported from the handheld probe to the computer, encoded, and then classified by their distances to dictionary of reference points. The clinician is presented with a report of each disease state probability, their distances from the reference points, and an attention heat map highlighting the regions of interest. }

\label{fig:FSL_Overview}
\end{figure}

FSL provides a framework for leveraging the knowledge domain of pretrained networks to novel tasks. In its basic form, high dimensional images are encoded into a metric feature space and then classified by their relations from learned reference points as shown in Figure \ref{fig:FSL_Overview}. FSL follows a long line of metric based learning with many recent works focusing on incorporating DNN. One early example is the Siamese network architecture \cite{Koch2015SiameseNN}. Further developed by the Matching network \cite{vinyals2017matching} where the DNN was trained to estimate the set-to-set probability. In 2017, Prototypical-Net \cite{snell2017prototypical} trained a DNN to directly generate a discriminative feature embedding space.

The application of ML in ultrasound diagnostics has been growing rapidly, but still limited when compared to other imaging modalities. These pioneering studies address such task as: tumor detection, fetal health monitoring, and cardiac monitoring \cite{2018_US_ML_Review,2019_US_ML_Review}. 

The onset of COVID-19 created a push for rapid pulmonary diagnostics. These developments have successfully utilized MRI, CT, and LUS images to detect COVID-19 characteristics in patients' lungs. To the best of our knowledge, there have only been four studies applying ML for LUS diagnostics \cite{auto_B_line,lung_abnorm_detect}, and only two focused on the detection of COVID-19\cite{born2020poCOVIDnet,born2020pocus}.

This work differs from the current SOTA in three major ways due to the direct consideration for clinical usage during development, primarily the ability to adapt to individual practices and provide information in such a way to be easily incorporated into the greater corpus of information used in the diagnostic process. These differences are: 1.) we present a full ML LUS diagnostic system; 2.) we incorporate a vocabulary based FSL pipeline enabling a significant reduction in training requirements; 3.) our system generates intuitively understandable distance based classifications.

% the significant reduction of training requirements enabling rapid adaptability to variable of machines and procedures seen in PoC environment. Through the introduction of a vocabulary based FSL with dictionary classification the proposed system establishes a framework

% The primary differences between our approach Born et al. \cite{born2020poCOVIDnet} in our few-shot approach. Born et al. \cite{born2020poCOVIDnet} trains a full DNN named POCOVID-Net based on a fully trained VGG-16 architecture. Instead, we implement a DNN pretrained on the ImageNet as a feature extractor, encoding each LUS image into its latent feature space. We then process this feature space using principal component analysis (PCA) dimension reduction and a k-means clusters derived vocabulary, inspired by Gidaris et al. \cite{gidaris2020learning}. Another differentiation of our algorithm is the Mahalanobis based classification, inspired by Bateni et al. \cite{bateni2020improved}. We use the processed feature space to generate a code book of class signatures which are then compared to query images. The query image is classified by its distance from the reference signatures giving comprehensible relative probabilities. In total, the proposed algorithm contains 329,216 parameters, which
% is a 97.7\% reduction from the 14,747,971 parameters used by the POCOVID-Net. 

%-------------------------------------------------------------------------
\section{Approach}
\begin{figure}[!h]
\begin{center}
\fbox{\includegraphics[width=.9\linewidth]{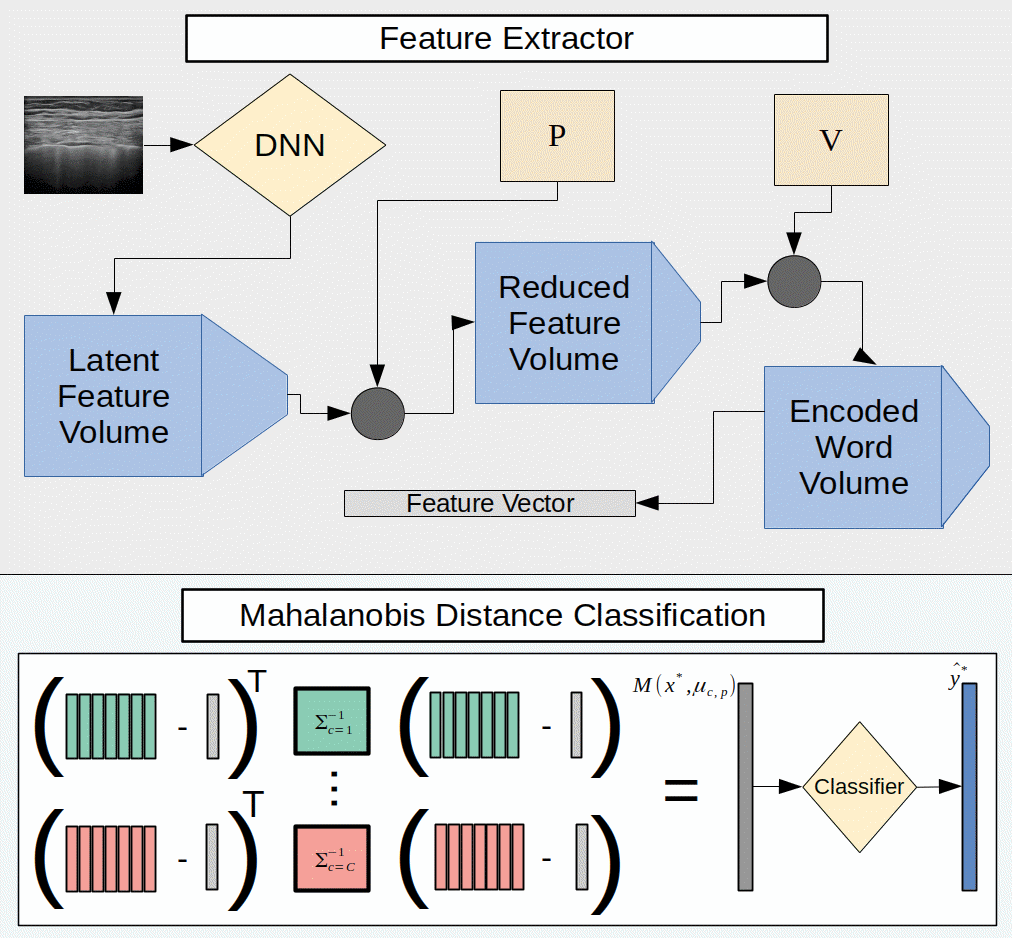}}
   
\end{center}
  \caption{\textbf{Algorithm Overview:} Few-shot visual classification is comprised of a feature extraction and classification process. The image features are extracted using the pretrained DNN; then PCA reduced and encoded with the learned vocabulary into a feature vector representation. The support set is used to generate a dictionary of class signatures consisting of the class centroid and covariance matrix. Query images are classified by their Mahalanobis distances to the dictionary signatures using linear discriminant analysis.}
\label{fig:algorithm_overview}

\end{figure}

This section presents the methodology behind the proposed approach. The flow chart of the algorithm is shown in Figure \ref{fig:algorithm_overview}. First the theory and problem formulation of FSL is presented. This is followed by an explanation of the feature extraction and classification processes and concludes with the algorithm training process.

\subsection{Problem Formulation}
Assume there is a dataset of \(N\) images, \(D={(x_i,y_i)}^{N}_{i=1}\), where  \(x_i \in \mathbb{R}^{d_{img}}\) with labels \(y_i \in {1,\dots,C}\).  The FSL approach uses a sub-sample \(D^{\tau}\subseteq D\) to create a set of support (reference) images, \(S^{\tau}={(x_i,y_i)}^{N^{\tau}}_{i=1}\), and a set of query images to be classified, \(Q^{\tau}={(x^{*}_{i},y^{*}_{i})}^{N^{*\tau}}_{i=1}\). FSL requires a few assumptions on the relationships between sets \(D, D^{\tau}, S^{\tau}, Q^{\tau}\). \(D^{\tau}\) must be a sub-sample of classes each with a specified number of samples, \(k\). \(S^{\tau}\) and \(Q^{\tau}\) must be split such that the classes in the query set are represented in the support set. The FSL objective is to find a function \(f(S^{\tau},Q^{\tau})\) that best estimates of the query set labels, \(y^{*}\), with best being defined as \(\mathbb{E}_{\tau}[\prod_{Q^{\tau}}p(y^{*}_{i}|f(x^{*}_{i},S^{\tau})]\). 

\subsection{Theory}
Deep neural image classification networks are trained to estimate the class probabilities of a given image using a final \(softmax()\) output layer. This process generates learned feature extracting filters corresponding to the most discriminative features of the training set. Assuming the learned filters are sufficiently generalized, a query image can be effectively encoded in the network's latent feature space.

\begin{equation}
p(y^*=c|x^*) = softmax(\Phi(x^*))
\end{equation}

The class probabilities can also be viewed as a Gaussian mixture model (GMM) of class characteristics in the latent manifold \cite{bateni2020improved}. This view looks at the image as an instance of characteristics from a GMM source. The DNN performs a series of linear kernel transformations. From the central limit theorem, it is known that the linear combination of Guassian distributions generates a Gaussian distribution. Therefore, the DNN embedded features can be viewed as a GMM produced from the GMM of visual characteristics with a class probability defined by:

\begin{equation}
p(y^*=c|x^*) = \frac{\pi_c \mathcal{N}_c(\mu_c,\Sigma^{\tau}_c)}{\Sigma_{c'}\pi'_{c}\mathcal{N}(\mu_{c'},\Sigma^{\tau}_c)}
\end{equation}

The proposed algorithm performs a series of linear transforms on the GMM of the latent manifold, preserving the mixture model throughout the process. The principle component analysis (PCA) calculates the dominant directions of the manifold through the eigenvectors of the covariance matrix. The result is a manifold oriented by decreasing variance and therefore decreasing entropy. Trimming the eigenvectors with the smallest eigenvalues reduces the GMM to the most informative distributions with the greatest entropy.

The k-means cluster vocabulary organizes the distribution structures into 'words' across classes according to the prominent feature clusters seen in the latent features of the support set. Interpreting the latent manifold with the calculated vocabulary combines distributions into semantically relevant features based on their similarity to the 'words' in the generated support set vocabulary.

The Mahalanobis distance transforms the GMM according to its relation to the learned class signatures. This process centers the GMM around the code word and scales it with the covariance matrix. Therefore, the Mahalanobis distances can also be considered as instances of a GMM and can be interpreted as the probability of the query image originating from the GMM of the reference class.

The final classification is completed using the linear discriminant analysis (LDA), which projects the class probabilities to the optimally discriminative space, maximizing interclass variance while minimizing the intraclass variance.

\paragraph{\textbf{Feature Extraction}}
The proposed algorithm extracts activation features from the latent space of a pretrained DNN, MobileNet \cite{howard2017mobilenets}. Let \(\Phi\) be the feature embedding function of the DNN. The feature embedding, \(a_i\), of the image \(x_i\) is generated by a forward pass through the network, shown in Equation \ref{eq:1}, where \(a_i\) are the activations within the networks latent space. The embedded features are then compressed to \(a'_i\) using Equation \ref{eq:2}. The latent features are reduced by PCA transform, \(P\), and interpreted by the vocabulary, \(V\), calculated from \(S^{\tau}\) using k-means clustering. The resulting mean vector of the features are then normalized, creating the image feature vector \(r_i\).

\begin{equation}\label{eq:1}
\Phi(x_i)=a_{i}
\end{equation}

\begin{equation}\label{eq:2}
   Pa^{T}_{i}V^{T}=a'^{T}_iV^{T}=r_i
\end{equation}

\paragraph{\textbf{Dictionary Generation}}
The class signatures (a.k.a. representative appearance models) of the dictionary are calculated from the image feature vectors, \(r(x)\) in the support set \(S^{\tau}\). The support set provides \(k\) examples for each class, giving us multiple representations per class \(r_{c,k}\). The co-variance of the class, \(\Sigma_c\), is calculated from \(S_c^{\tau}\). Class Sub-representations are calculated from the k-means of the \(r_{c}\) manifold, producing \(p\) clusters. The hierarchical code word representations are generated from the centroids of each cluster, \(\mu_{c,p}\). 

\paragraph{\textbf{Classification}}
The class of a query image is predicted by the distances of its signature to those in the dictionary. The distances are calculated using Mahalanobis distances from the class signatures, \(\mu_{c,p}\).
\begin{equation}
M(x^*,\mu_{c,p})=\frac{1}{2} (x^*-\mu_{c,p})^T \Sigma^{-1}_c (x^*-\mu_{c,p})
\end{equation}

The final classification decision is made by linear discriminant analysis (LDA) of the query image distances to each dictionary signature.

\paragraph{\textbf{Training}}
The proposed approach is unique in that it requires no DNN training. Instead, a series of linear transforms is trained on small sample of reference images and project the MobileNet DNN embedded features to a optimally discriminative space. These transforms include PCA reduction, k-means vocabulary, dictionary, and LDA separation. The PCA is pretrained on an unlabeled random sub-sample of \(D\), serving as a general context transform of the DNN response to a lower dimensional space, trimming low activation neurons. The k-means, dictionary, and LDA are trained on the sub-sampled support set of reference images, \(S^{\tau}\). 

\subsection{COVID-19 POCUS Dataset}
Performance evaluations are conducted on the COVID-19 POCUS Dataset \cite{born2020pocus}, the largest publicly available of its type, comprising PoC LUS images from COVID-19, pneumonia, and healthy patients. The dataset is split into LUS clips produced by linear and convex probes. The clips were collected from several sources, the primary being: grepmed.com, thepocusatlas.com, butterflynetwork.com and radiopaedia.org. The dataset is heterogeneous, originating from varying institutions and devices. The data is unidentified and no additional meta data, such as vitals or demographics, are provided. All image annotations are verified by medical professionals. In total, the linear clips contain images comprised of 1,457 normal, 315 pneumonia, 445 COVID frames. The convex LUS clips contain images comprised of 11,646 normal, 4,585 pneumonia, 8,188 COVID frames. The proposed algorithm was evaluated on test data randomly selected and sequestered using a 20\% split. Each image was normalized and resized to (224,224).

\section{Experiments}
The objective of the experiment is to analyze the training requirements and classification performance of the system. This is done by running an analysis of the algorithm's classification performance with varying number of reference images from 8 to 64 for each class. Three binary classification scenarios are considered: healthy v. COVID-19, healthy v. pneumonia, and pneumonia v. COVID-19. The algorithm was implemented in Python on a Linux OS with open-source libraries.  All experiments were ran on a Intel(R) Core(TM) i5-8600K CPU with 16 GB of RAM. The longest experimental case (using 64 samples) took 15 seconds to process. All evaluation metrics are calculated over 10 trials each containing randomly selected training and test sets.

\subsection{Results}
The experimental performances was evaluated using receiver operating characteristic curves (ROC) which shows the system's sensitivity over its selectivity. Note the ROC curves are plotted using 1-Specificity for easier reading. Only results for linear ultrasound images are shown due to limited space.

\begin{figure}[!htbp]
\begin{center}
%\fbox{\rule{0pt}{2in} \rule{0.8\linewidth}{0pt}}
\includegraphics[width=.8\linewidth]{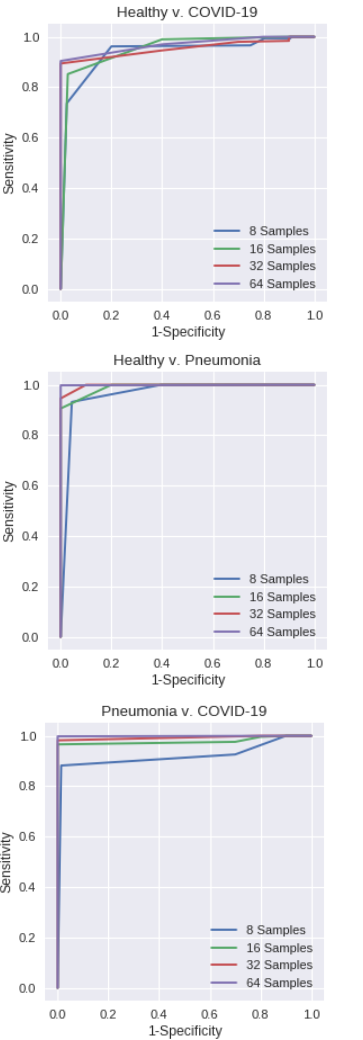}
\end{center}
\caption{\textbf{Linear US ROC Curves:} These plots show the ROC curves of the three classification scenarios: healthy v. COVID-19, healthy v. pneumonia, and pneumonia v. COVID-19. These plots contain curves for each considered number of training samples. Note these plots use 1-specificity.  }
\label{fig:roc_curves}
\end{figure}

Figure \ref{fig:roc_curves} shows the mean ROC curves for each experimental case. The plots are organized first by classification scenario, (healthy v. COVID-19, healthy v. pneumonia, or pneumonia v. COVID-19), and then by number of training samples per class, (8,16,32,64). The strongest trend is seen in the increase in specificity with number of training examples. A performance saturation is seen at 64 samples for all scenarios. The highest performance is seen in the healthy v. pneumonia case achieving a high sensitivity with just 8 training samples. This was followed by the pneumonia v. COVID-19 and then healthy v. COVID-19. These results show that detecting COVID-19 is a more challenging task than detecting pneumonia, but still be achieved with 64 samples per class.

\begin{figure}[!ht]
\begin{center}
%\fbox{\rule{0pt}{2in} \rule{0.8\linewidth}{0pt}}
\includegraphics[width=.9\linewidth]{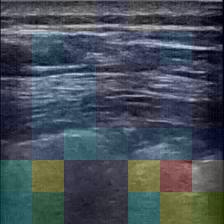}
\end{center}
\caption{\textbf{Attention Heat Maps:} This is an example image of a COVID-19 attention heat map on a COVID-19 positive ultrasound image. The image is sampled in a grid of patches with each patch colored by its relative distance to the learned class signatures.}\label{fig:heat_maps}
\end{figure}

In the pursuit of increased decision understanding an attention heat map to highlight the image regions that correspond to the algorithm's disease state decision is generated by the system. Only qualitative assessment is possible due to a lack of segementation annotation. Figure \ref{fig:heat_maps} shows an ultrasound image of a COVID-19 infected lung with the attention heat map. The image is sampled in a grid of image tiles. The distance of each tile to the learned COVID-19 signature is denoted by its color with red being the highest. This heat map highlights a sub-pleural consolidation.

\section{Discussion}
The purpose of these experiments is to asses the potential effectiveness of the COVID-19 ultrasound diagnostic system evaluated by its ability to accurately predict disease state and to efficiently learn from limited samples. In medical decision making, the risks of type one and two errors must be considered. The ROC curves display the performance trade-offs for higher sensitivity and specificity. The results of the experiments show that the algorithm is capable of reliably detecting COVID-19 symptoms in the lungs. the results also show that the algorithm is capable of reliably distinguishing COVID-19 symptoms from pneumonia.

The results of the experiments also demonstrate a significant reduction in training requirements with the capability of learning disease models with as few as 8 training samples per class in some scenarios. A high classification performance was seen in all scenarios with 64 samples per class. This capability opens the door for clinicians to adapt the algorithm to their environmental factors, such as differences in patient demographics, equipment, and operators.

The value of the algorithm's diagnostic performance is dependent on its ability to be incorporated into the larger clinical diagnostic process. This requires that the algorithm's diagnostic predictions be presented in an intuitively understandable manner. Qualitative assessment of the attention heat maps demonstrates the capability of highlighting relevant regions of interest. The combination of the high prediction performances and the intuitive displays of the system delivers the aide of deep learning in a clinically viable way.

\section{Conclusion}
Rapid, accurate, and inexpensive COVID-19 detection is critically needed. This paper presents the adaptive PoC ultrasound COVID-19 diagnostic system based on the novel FSL visual classification algorithm. The system was designed with a specific focus on its incorporation into the clinical diagnostic process, requiring understandable outputs, adaptability, and reliability. The system takes less then 15 seconds to train on an Intel(R) Core(TM) i5-8600K CPU. The generated disease state models are compact each requiring less than 1 MB of memory. The distance based classification provide intuitive interpretation of the system's predictions. The attention heat maps highlight the regions of the ultrasound images that are most responsible for its classification. The results show that the system is highly capable of accurately diagnosing COVID-19 and pneumonia disease states with as few as 64 training images per disease.

% The results of this study demonstrate the feasibility of using FSL for the detection of Covid-19 and pneumonia using PoC LUS imaging. The proposed algorithm has a 97.7\% parameter reduction compared to the state-of-the-art POCOVID-Net while surpassing its performance. The fact that such a high performance was achieved with such limited training data demonstrated the adaptability of our algorithm, opening the door to wider healthcare applicability. The methodology we adopted in this study makes it uniquely possible to train algorithms using small data sets to identify serious medical conditions and direct timely treatment, which can potentially transform the standard of care in PoC settings by enabling non-experts to use an automated diagnostic tool to identify the life-threatening conditions.

{\small
\bibliographystyle{ieeetr}
\bibliography{COVID_bib}
}

\end{document}